\documentclass[twocolumns]{mn2e}

\usepackage{xcolor,soul}
\input psfig.sty

\def\T1{\ {$T_1$}\ }
\def\MT1{\ {$M_{T_1}$}\ }
\def\ct1{\ {$(C-T_1)$}\ }
\def\CT10{\ {$(C-T_1)_0$}\ }

\def\VI0{\ {$(V-I)_0$}\ }

\def\2cd{\ {two-color diagram}\ }

\def\sf{\ {specific frequency}\ }
\def\ell{\ {elliptical}\ }

\def\gtsim{\ {\raise-0.5ex\hbox{$\buildrel>\over\sim$}}\ }
\def\ltsim{\ {\raise-0.5ex\hbox{$\buildrel<\over\sim$}}\ }

\begin{document}

\title[LMC star clusters]{A new Extended Main Sequence Turnoff star cluster in the Large Magellanic Cloud}

\author[A.E. Piatti]{Andr\'es E. Piatti\thanks{E-mail: 
andres@iafe.uba.ar}\\
Instituto de Astronom\'{\i}a y F\'{\i}sica del Espacio, CC 67, Suc. 
28, 1428, Ciudad de Buenos Aires, Argentina\\
}

\maketitle

\sethlcolor{green}
\begin{abstract}
We present results on the age and metallicity estimates of the poorly studied 
LMC cluster SL\,529, from CCD SDSS $gr$ photometry obtained at the Gemini South telescope
with the GMOS attached. The cluster MSTO region possesses an
extended structure, with an age spread ($\sim$ 0.5 Gyr) bigger than the mean age width of known EMSTO LMC clusters.
We report for the first time a mean cluster age of 2.25 Gyr and a mean
cluster metallicity of Z=0.004,  which place it as the most 
metal-poor and oldest cluster in the EMSTO LMC cluster group.
In addition, the cluster RC appears to
be formed by two concentrations of stars - although it is not clear whether this feature can be caused, in
part, by binary interactions and mergers -,
whereas the cluster core radius of 4.2 pc is in excellent agreement with those determined for 
the previously 12 known EMSTO LMC clusters.
\end{abstract}

\begin{keywords}
techniques: photometric -- galaxies: individual: LMC -- Magellanic 
Clouds -- galaxies: star clusters. 
\end{keywords}

\sethlcolor{green}
\section{Introduction}

Up to date, only 12 Large Magellanic Cloud (LMC) star clusters have been uncovered to possess
Extended Main Sequence Turnoffs (EMSTOs), namely: NGC\,1751, 1783, 1806, 1846, 1852, 1917, 1987,  
2108, 2154, 2173, Hodge\,7, and SL\,862 (see, e.g., Bertelli et al. 2003, Milone et al. 2009). The age range of this cluster sample 
is $\sim$ 1-2 Gyr, while a roughly constant metal content of [Fe/H] = -0.4 dex has been reported
for them Mackey \& Broby Nielsen (2007), Mackey et al. (2008), Goudfrooij et al. (2009), Girardi et al. (2009). These age and metallicity ranges coincide with those of an important bursting cluster
formation epoch that took place in the LMC $\sim$ 2 Gyr ago, possible due the tidal interaction
between both Magellanic Clouds and, perhaps, also the Milky Way. Particularly, Piatti \shortcite{p11} 
found that the number of studied clusters with ages between 1 and 3 Gyr now doubles that of the population of clusters formed at the bursting epoch, thus increasing the probability that more EMSTO star clusters be identified. 
However, from a total of some 70 star clusters with age estimates within our age range of interest, the
list of known EMSTO clusters has remained unchanged. The  relatively small number 
of known EMSTO LMC star 
clusters is even more noticeable when considering the 2268 star clusters cataloged
in the LMC (Bonatto \& Bica, 2010), thus representing $\sim$ 3\% of the estimated LMC population of clusters formed at the bursting epoch.

In this paper, we report for the first time age and metallicity estimates for SL\,529
(R.A.= 5$^h$31$^m$06$^s$, Dec.= -63$\degr$32$\arcmin$23$\arcsec$, J2000), a poorly studied 
LMC star cluster located in the outer disk of the galaxy. The results show that this cluster
belongs to the handful of EMSTO clusters in the LMC. The impact of this finding would appear to be twofold:
first, we actually found a new LMC cluster formed at the begining of the bursting formation epoch
which exhibits the EMSTO phenomenon. To this respect, we do show not only evidence about the age spread in the 
cluster MSTO, but also that the magnitude of the age spread exhibited by its stellar populations is correlated 
with the cluster core radius \cite{ketal11}. Additionally, we identified 
two concentrations of stars in Red Clump (RC), although
it is not clear whether this feature can be caused, in
part, by binary interactions and mergers \cite{yetal11}. 

Second, SL\,529 resulted to be the oldest (age $\sim$ 
2.25 Gyr) and the most metal-poor EMSTO LMC cluster ([Fe/H] $\sim$ 
-0.7 dex). The cluster age is close to the limit estimated by Keller et al. \shortcite[age $\sim$ 2.3 Gyr]{ketal11} from which
the age spread of the cluster stellar populations represents a diminishing fraction of the cluster age, 
so that the multiple populations become increasingly harder to resolve photometrically. On the other
hand, the resulting cluster metallicity is lower than any known 1-3 Gyr old LMC cluster 
\cite{pg12}, making it a interesting case to constrain EMSTO models at different overall 
chemical abundance levels. 

This paper is organized as follows. In Section 2 we describe
the data collected, the reduction procedures performed, and the
subsequent photometry standardization. Section 3 deals with the analysis of the data,
from which cluster age, metallicity, reddening and structural parameters have been derived.
In this Section we provide with observational evidence about the status of SL\,529 as an
EMSTO LMC cluster. Finally, we summarize our results in Section 4.

\section{Data handling}

Based on data obtained from the Gemini Science Archive, 
we collected CCD SDSS $gr$ (Fukugita et al., 1996) images centred on 26 LMC clusters
(GS-2010B-Q-74, PI: Pessev) along with observations of standard
fields and calibration frames (zero,
sky-flat, dome-flat). The data were obtained at the Gemini South telescope with the 
Gemini Multi-Object Spectrograph (GMOS) attached (scale = 0.146 pixel$^{-1}$). 
The data consist in 2 exposures of 30 s in $g$ and 2 exposures of 15 s in $r$ 
for each cluster under seeing conditions
better than 0$\arcsec$.6 and with airmass of 1.25-1.40. 
Nine Gemini Observatory standard fields 
were observed along the 5 cluster observing nights, for which 
%and, in some cases, at different airmass during the same night, 
2 exposures of 5 s per filter and 
airmass in the range $\sim$ 1.0-2.0 were obtained.

The data reduction followed the procedures documented by
the Gemini Observatory 
webpage\footnote{http://www.gemini.edu/sciops/instruments/?q=sciops/instruments} and  the IRAF\footnote{ 
IRAF is distributed by the National 
Optical Astronomy Observatories, which is operated by the Association of 
Universities for Research in Astronomy, Inc., under contract with the National 
Science Foundation.}.{\sc gmos} package. We processed a total
of 118 images (standard star and cluster fields) by performing overscan, trimming corrections,
bias subtraction, flattened all data images, etc., once the calibration
frames (zeros, sky- and dome-flats, etc.) were
properly combined. The final field of
view of the images resulted to be $\sim$ 5$\arcmin$.6 $\times$ 5$\arcmin$.6.

Depending on the night, nearly 30-50 independent magnitude measures of standard stars
were derived per filter using
the {\sc apphot} task within IRAF, in order to secure the transformation
from the instrumental to the SDSS $gr$ standard system. The relationships between
instrumental and standard magnitudes were obtained by fitting
the following equations:

\begin{equation}
g = g_1 + g_{std} + 0.18\times X_g + g_2\times (g-r)_{std}
\end{equation}

\begin{equation}
r = r_1 + r_{std} + 0.10\times X_r + r_2\times (g-r)_{std}
\end{equation}

\noindent where $g_i$, and $r_i$ (i=1,2) are the fitted coefficients, and
$X$ represents the effective airmass. We solved the transformation equations with the {\sc fitparams}
task in IRAF and found mean colour
terms of -0.058 in $g$, -0.049 in $r$. The rms errors from
the transformation to the standard system were 0.015 mag for $g$ and 0.023 for $r$, respectively, 
indicating an excellent photometric quality.

The stellar photometry was performed using the starfinding
and point-spread-function (PSF) fitting routines in
the {\sc daophot/allstar} suite of programs \cite{setal90}.
For each image, a quadratically varying PSF was derived by
fitting $\sim$ 60 stars, once the neighbours were eliminated using a
preliminary PSF derived from the brightest, least contaminated
$\sim$ 20 stars. Both groups of PSF stars were interactively selected.
We then used the {\sc allstar} program to apply the resulting PSF
to the identified stellar objects and to create a subtracted image
which was used to find and measure magnitudes of additional
fainter stars. This procedure was repeated three times
for each frame. After deriving the photometry for all detected objects in
each filter, a cut was made on the basis of the parameters
returned by {\sc daophot}. Only objects with $\chi$ $<$2, photometric error less than 
2$\sigma$ above the mean error at a given
magnitude, and $|$SHARP$|$ $<$ 0.5 were kept in each filter.

Finally, we standardized the resulting instrumental magnitudes and combined all the
independent measurements using the stand-alone {\sc daomatch} and {\sc daomaster} programs, 
which were kindly provided by Peter Stetson. 
The final information consists of a running number per star, its $x$ and $y$ coordinates,
the mean $g$ magnitudes and $g-r$ colours, the
standard errors $\sigma$($g$) and $\sigma$($g-r$), and the number $n$ of averaged magnitudes and 
colours. 
%Whenever $n$ is equals to one, we list the observational errors provided by {\sc daophot}.
A portion of Table 1, which gives this information for a total of 5454 stars in the field
of SL\,529, is shown here for guidance regarding its form and content, while Fig. 1 shows
the obtained magnitude and colour errors. The
whole content of Table 1 is available in the online version of the
journal. The photometric information gathered for the remaining 25 clusters will be presented 
in a forthcoming paper.

\begin{figure}
\centerline{\psfig{figure=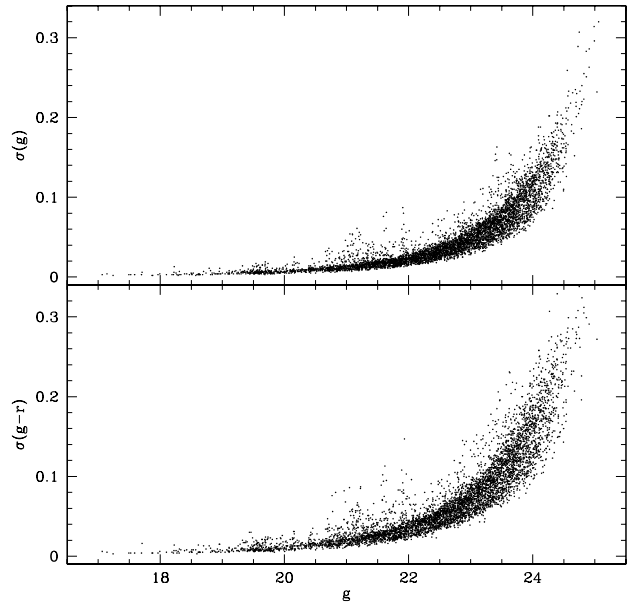,width=84mm}}
\caption{Magnitude and colour errors as a function of $g$ for stars which passed all the 
required photometric restrictions.}
\label{fig1}
\end{figure}

\section{Data analysis}

The cluster centre [($x_c$,$y_c$)=(1520$\pm$10,1100$\pm$10) pixels] was estimated using the {\sc ngaussfit} 
task within the IRAF.{\sc stsdas} package. 
We then constructed the cluster radial profile depicted in Fig. 2 (right panel), which served us to adopt the
cluster radius ($r_{cls}$) -defined as the distance from the cluster centre where the stellar density
profile intersects the background level- to perform circular extractions in the Colour-Magnitude Diagram (CMD). 
We also fitted a King \shortcite{k62} model to the normalized radial stellar density profile using the expression:

\begin{equation}
N/N_o = ({\frac{1}{\sqrt{1+(r/r_c)^2}} - \frac{1}{\sqrt{1 + (r_t/r_c)^2}}})^2 + bkg
\end{equation}

\noindent where $N_o$ is the central stellar density, and $r_c$ and $r_t$ are the core and tidal radii, respectively.
$bkg$ represents the background level. The values for $r_c$, $r_{cls}$, and $r_t$ turned out to be (110$\pm$10) 
pixels, (450$\pm$50) pixels, and (1300$\pm$100) pixels, respectively. Circles centred on the cluster with radii
$r_c$ and $r_{cls}$ are drawn in the left panel of Fig. 2, and they are also indicated with vertical lines in the 
right panel of the figure.

\begin{figure}
\centerline{\psfig{figure=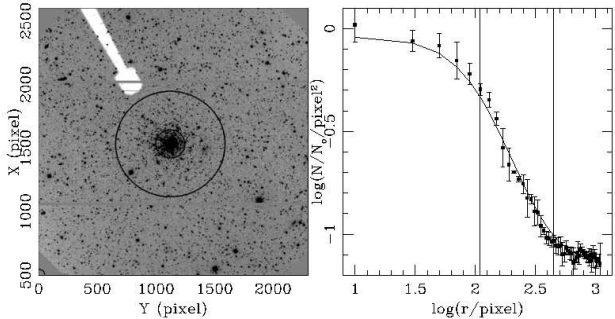,width=84mm}}
\caption{{\it Left panel:} 15 s exposure $r$ image centred on SL\,529. The circles corresponding to $r_c$ and
$r_{cls}$ are also drawn. {\it Right panel:} The cluster stellar radial profile (filled box) with the
derived errorbars, and the best-fitted King (1962) model superimposed. The vertical lines represent
$r_c$ and $r_{cls}$.}
\label{fig2}
\end{figure}

We built two CMDs centred on the cluster, one for a circular extraction of radius $r_{cls}$, and another
with an inner radius 2$\times$$r_{cls}$ and equal area as for the cluster. The resulting cluster and field
CMDs are shown in Fig. 3. As can be seen, the cluster MSTO appears noticeably extended, the cluster RC shows a
dual structure, and the cluster Red Giant Branch (RGB) resembles that of a single overall metallicity. All these
features suggest that SL\,529 could be included from now on in the list of known EMSTO LMC clusters.

\begin{figure}
\centerline{\psfig{figure=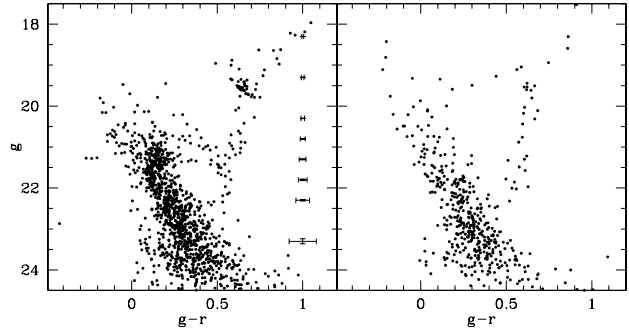,width=84mm}}
\caption{Colour-Magnitude diagrams of SL\,529 (left) and of its surrounding field (right) for an
equal cluster area. The magnitude and colour errorbars are also indicated.}
\label{fig3}
\end{figure}

In order to estimate the mean cluster age, we adopted a distance 
modulus of $(m-M)_o$ = 18.50 $\pm$ 0.10 \cite{getal10}. 
The reddening value was taken from the Burstein \& Heiles \shortcite{bh82}, and 
Schlegel, Finkbeiner, \&  Davis \shortcite{sfd98} extinction maps, adopting a weighted average value 
of $E(B-V)$ = 0.04 $\pm$ 0.01, since both the small reddening values as well as their errors suggest 
that no differential reddening would affect the cluster. 
We adopted $R$ = $A_V$/$E(B-V)$ = 3.1 to convert the colour excess to 
the extinction, and
 used the equations $A_g$/$A_V$ = 1.199 and $A_r$/$A_V$ = 0.858 \cite{f99} to evaluate the total extinctions 
in $A_g$ and $A_r$. Finally, we used $E(g - r)$/$A_V$ = 0.341 for the selective extinction in the SDSS system.

When we chose subsets of isochrones for different Z metallcity values to evaluate the 
metallicity effect in the cluster age, we preferred to follow the general rule of 
starting  without assuming any prearranged metallicity. Instead, we adopted chemical compositions 
of Z = 0.002, 0.004 and 0.008 ([FeH] = -1.0, -0.7, and -0.4, respectively) for the isochrone sets which 
cover the metallicity range of most of the intermediate-age LMC clusters studied in detail so far (Piatti \& Geisler 2012). We then selected a set of isochrones of Marigo et al. (2008) and superimposed them on the cluster CMD, once they were 
properly shifted by the corresponding $E(B-V)$ colour excesses and by the
adopted LMC apparent distance 
modulus. 
%In the matching procedure, we used isochrones for each metallicity 
%level, ranging from slightly younger than the derived cluster age to slightly older. 
Finally, 
we adopted as the cluster age the one corresponding to the isochrone 
of age = 2.25 Gyr and Z = 0.004 which best reproduces the cluster 
main features in the CMD. 
%The presence of RCs and/or RGBs in some cluster CMDs made the fitting procedure 
%easier. We noted, however, that the theoretically computed bluest stage during the He-burning core phase 
%is redder than the observed RC in the CMDs of some clusters, a behaviour already 
%detected in other studies of Galactic and Magellanic Cloud clusters (e.g., Geisler et al. 2003; Piatti et 
%al. 2004a; Piatti et al. 2004b). A similar outcome was found from the fitting of isochrones in the 
%$M_V$ vs $(V-I)_o$ diagram \cite[among others]{petal03b,petal03c}. 
Fig. 4  shows the result of the fitting, wherein we plotted the isochrone 
of the adopted cluster age and two additional isochrones bracketing the derived age.  The bracketing isochrones (age = 2.0 and 2.5 Gyr) were fitted by taking into account the observed spread in the cluster MSTO.
We would like to note that the estimated
cluster age and metallicity place it as the most metal-poor and oldest EMSTO LMC cluster.

\begin{figure}
\centerline{\psfig{figure=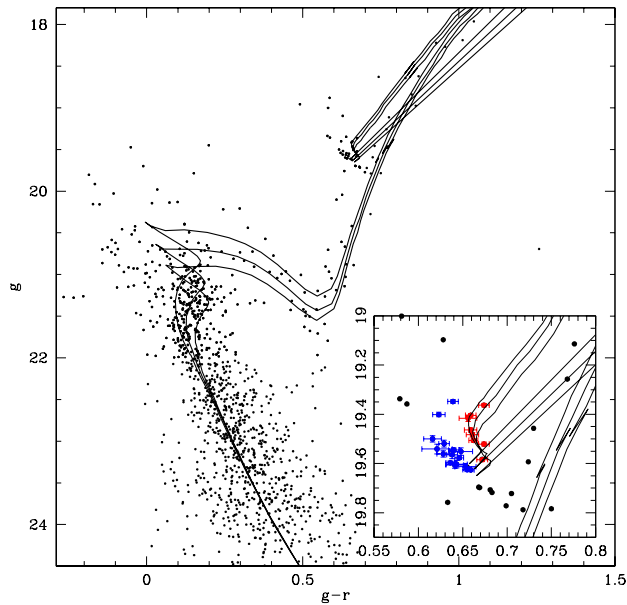,width=84mm}}
\caption{$g$ versus $g-r$ cluster CMD with isochrones of Marigo et al. (2008) for $t$ = 2.0, 2.25, and 2.5 Gyr and Z = 0.004 superimposed. The inner panel depicts an enlargement of the RC region, wherein two groups
of RC stars have been represented with blue and red circles.}
\label{fig4}
\end{figure}

In order to quantify the MSTO extension we counted the number of stars (N) located within the rectangle
drawn in Fig. 5, using as independent variable the coordinate (X) along the long axis, according
to the precepts outlined by Goudfrooij et al. (2011, see their Fig. 14). 
The resulting field subtracted histogram is depicted in Fig. 6, which clearly exhibits an extended distribution.
When building the histogram we have taken into account the effects caused by using different binning as well
as the point errors, which cause the points have the chance to fall outside their own bin if they do not
fall in the bin centre (Piatti \& Geisler 2012); thus producing an intrinsic distribution.
%We then fitted two Gaussian functions to Fig. 5 using {\sc ngaussfit} and, once the X coordinate is converted into 
%age, we derived age$_1$ = \hl{(2.85 $\pm$ 0.20)} Gyr and $FWHM$$_1$ = \hl{(0.50 $\pm$ 0.05)} Gyr  and age$_2$ = 
%\hl{(3.35 $\pm$ 0.05)} and $FWHM$$_2$ = \hl{(0.30 $\pm$ 0.05)} Gyr. \hl{Notice that, although both peaks results 
%distinguishable 
%($|$age$_2$-age$_1$$|$ $>$ ($FWHM$$_1$+$FWHM$$_2$)/2),
Fig. 6 also compares this resulting distribution with those of Goudfrooij et al. 
(2011) for the known EMSTO LMC clusters NGC\,1751 (blue line), NGC\,1783
(red line), NGC\,1806 (yellow line), NGC\,1987 (magenta line), and
NGC\,2108 (cyan line), previously normalized to the unity and shifted
in order to superimpose the youngest ends. As can be seen, the most important result is that we
confirm the existence of an EMSTO ($\sim$ 0.5 Gyr) for SL\,529, which is bigger than the mean age width of known EMSTO clusters 
\cite{metal09,getal11}.

The magnitude of the age spread exhibited by the stellar population of an EMSTO cluster
is correlated with the cluster core radius \cite{ketal11}. The 12 known EMSTO clusters do have core radii
bigger than $r_c$ $\sim$ 3.7 pc. In the case of SL\,529, we used the derived cluster core radius (in pixels) and
the recently estimated LMC distance of (53.5 $\pm$ 0.4) kpc \cite{hetal12} to obtain $r_c$ = (4.20 $\pm$ 0.35) pc. When entering this core radius and the cluster age in their Fig. 5, SL\,529 falls into the visibility window of EMSTO clusters, for which different authors have argued that they are witnesses of 
different/prolonged stellar formation
epochs (Keller et al., 2012, and references therein).

Indeed, Mackey et al. (2008) built synthetic CMDs to investigate the role played
by unresolved binary stars around the MSTO region of three LMC clusters (NGC\,1783, 1806, and 1846).
They found that the resulting CMDs
for a single population -corresponding to an isochrone with the age and metallicity of the respective cluster- do not reproduce alone the EMSTOs seen in the studied clusters, although they included a significant fraction of unresolved binaries. However, when considering
two stellar populations, beside a fraction of unresolved binaries, the resulting CMDs reproduce
better the observed ones. The age spread of these three clusters turned out to be $\sim$ 0.3-0.4 Gyr. 
Similarly, Goudfrooij et al. (2011) argued that the impact of interacting binaries as
suggested by Yang et al. (2011) in the CMDs of 7 EMSTOs LMC clusters studied by them 
($\Delta($age$_{MSTO}$) = 0.35$\pm$0.10 Gyr) is insignificant. They concluded that
the placement in their CMDs of these binaries stars most likely correspond to that of LMC field stars.
They are a low stellar density population in the CMD towards brighter
magnitudes and bluer colours than the cluster MSTO. Milone et al. (2009), among others, also claimed that 
taking into account the presence of binary
stars alone in intermediate-age LMC clusters is not enough to reproduce their 
observed EMSTOs ($\Delta$(age) $\ga$ 0.3 Gyr). According to these results and bearing in mind the 
age spread found for the SL\,529's MSTO, we also infer that more than one stellar population 
would appear to be required to explain the morphology of the MSTO region. Recently, Li et al. (2012)
have shown that most CMDs, including extended or multiple TOs, can be explained using simple stellar populations including both binary and stellar rotation effects, or composite populations with two components.

Dual red clumps (RCs) have also been observed in some star clusters with or
without EMSTOs (Girardi et al. 2010). However, while Girardi et al. (2009) suggested that 
the origin for such a structure is related to an age spread, Yang et al. (2011)
found that part of the observed dual RCs may be the result of binary interactions and mergers.
In order to highlight the RC structure of SL\,529 we plotted in Fig. 4 an enlargement of the RC
region, where 
two groups of stars have been represented with blue and red circles. The errorbar represents the star photometric errors. Notice that all the stars are located within $r_{cls}$ 
(see Fig. 1), and a slightly more populated picture would be obtained if we included 
cluster red giants spread within 1.5$r_{cls}$. The additional RC stars would have reached the outer cluster region due to the dynamical relaxation (Santiago 2009). Particularly, some four
stars would be added around ($g-r$, $g$) $\sim$  (0.6, 19.8) mag. We think that this handful of stars 
does not constitute a clear evidence of the existence of a secondary RC as suggested by
Girardi et al. (2009). On the other hand,
the RC interacting binaries modeled by Yang et al. (2011, see red points in their Fig. 1) would appear to cover the RC region as 
well as to reach fainter magnitudes ($\Delta$$V$ $<$ 0.5 mag) and bluer colours ($\Delta$($V-I$) 
$<$ 0.1 mag). Unfortunately, they did not present CMDs for ages larger than 1.8 Gyr,
although they mentioned that interacting binaries appear mainly in clusters with 1.2 Gyr $<$ age 
$<$ 3.0 Gyr. Despite the lack of a direct transformation of our ($g-r$,$g$) CMD to
their ($V-I$, $V$) one, the overall appearance of both RCs would seem to be similar. Further
spectroscopic studies are still needed in order to obtain more reliable membership status and to confirm whether the cluster RC actually contains a binary star population.
We would like to point out, however, that the age sequence in the
single stellar population models of Yang et al. runs perpendicular to that
of the Marigo et al.'s isochrones.

\begin{figure}
\centerline{\psfig{figure=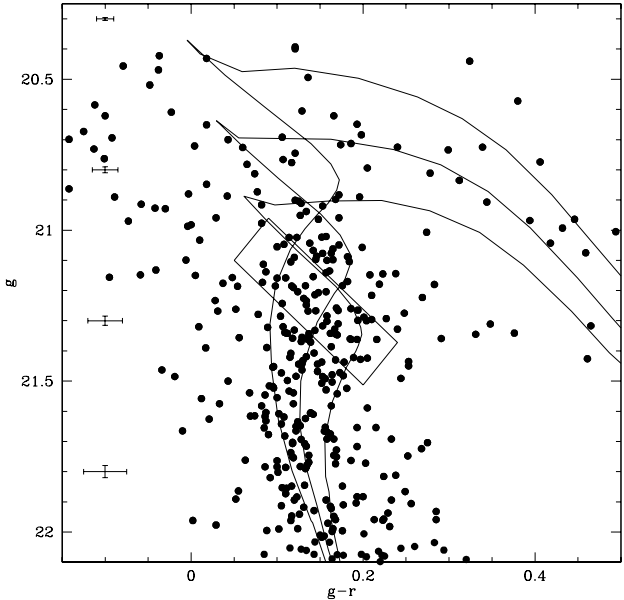,width=84mm}}
\caption{Enlargement of the CMD of SL\,529, focusing on the EMSTO region, with isochrones of Marigo et al.
(2008) for ages of 2.0, 2.25, and 2.5 Gyr, overplotted. 
A rectangle scanning the MSTO region is also superimposed.}
\label{fig5}
\end{figure}

\begin{figure}
\centerline{\psfig{figure=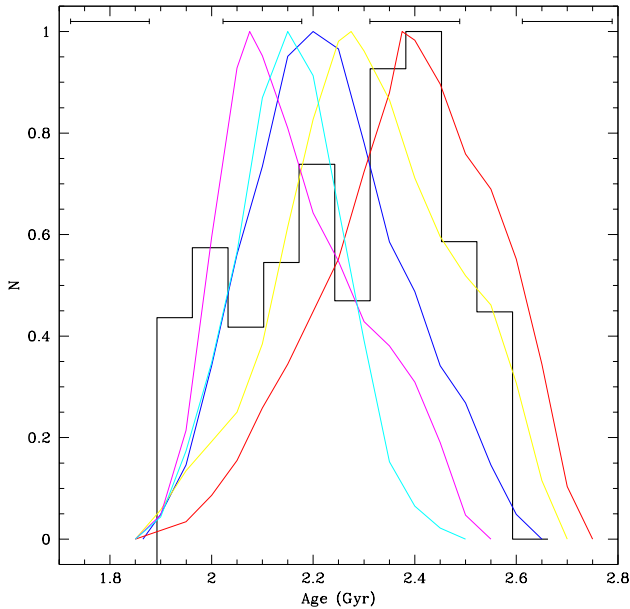,width=84mm}}
\caption{Number of stars counted along the major axis of the rectangle drawn in Fig. 5. The normalized MSTO distributions for NGC\,1751 (blue line), NGC\,1783 (red line), NGC\,1806 (yellow line), NGC\,1987 (magenta line), and
NGC\,2108 (cyan line) from Goudfrooij et al. (2011), are overplotted. 
The age errorbars spanning 2$\sigma$ are also illustrated.}
\label{fig6}
\end{figure}

\section{Summary}

In this study we present for the first time CCD SDSS $gr$ photometry
of stars in the field of the poorly studied LMC cluster SL\,529. 
The data were obtained at the Gemini South telescope with the GMOS attached
under high quality photometric conditions. We are confident
that the photometric data yield
accurate morphology and position of the main cluster features in the CMD.
The analysis of the cluster MSTO region shows that SL\,529 possesses an
extended MSTO of $\sim$ 0.5 Gyr.
We thus report that SL\,529 belongs to the group of the EMSTO LMC clusters.
We estimated the cluster age and metallicity by fitting theoretical isochrones to its CMD, previously shifted by its $E(B-V)$ colour excess and
apparent distance modulus. The isochrone which best reproduces the cluster
features turns out ot be that of 2 .25 Gyr and Z=0.004, with an observed
age spread of $\pm$ 0.25 Gyr. The estimated
cluster age and metallicity place it as the most metal-poor and oldest EMSTO LMC cluster.

In addition, we also found that the cluster RC appears to
be formed by two concentrations of stars, although it is not clear whether this feature can be caused, in
part, by binary interactions and mergers.
Finally, we fitted a King profile to the cluster stellar density radial distribution and
obtained a core radius of (4.20 $\pm$ 0.35) pc,  if a LMC distance of (53.5 $\pm$ 0.4) kpc 
is adopted. This value reinforces the status of SL\,529 as a EMSTO cluster, since it is
in excellent agreement with those determined for the 12 known EMSTO LMC
clusters.

\section*{Acknowledgements}

We greatly appreciate the comments and suggestions raised by the
reviewer which helped me to improve the manuscript.
This work was partially supported by the Argentinian institutions CONICET and
Agencia Nacional de Promoci\'on Cient\'{\i}fica y Tecnol\'ogica (ANPCyT). 

\sethlcolor{green}

\begin{table}
\caption{CCD SDSS $gr$ data of stars in the field of SL\,529.}
\begin{tabular}{@{}lccccccc}\hline
ID & $x$  & $y$ & $g$ & $\sigma$$(g)$ & $g-r$ & $\sigma$$(g-r)$ & $n$ \\
     & (pixel) & (pixel) & (mag) & (mag) & (mag) & (mag) & \\\hline
   ... & ...      & ...     & ...    & ...   & ...    & ... & ...   \\ 
     19 & 787.031 &1523.905 &  18.004 &   0.003 &   0.772 &   0.004 &  2\\
     20 &1198.930 & 767.067 &  18.007 &   0.007 &  -0.324 &   0.009 &  2\\
     21 &1651.733 &1353.525 &  17.968 &   0.003 &   0.866 &   0.004 &  2\\
   ... & ...      & ...     & ...    & ...   & ...    & ...  & ...  \\ 
\hline
\end{tabular}
\end{table}

\end{document}